\documentclass[twocolumn]{aastex631}

\newcommand\kms{km$\,$s$^{-1}$}
\newcommand\Msol{M$_{\odot}$}

\newcommand{\hi}{H\,{\sc i}}
\newcommand{\hii}{H\,{\sc ii}}

\graphicspath{{./}{figures/}}

\begin{document}

\title{Corvus~A: A low-mass, isolated galaxy at 3.5~Mpc}

\correspondingauthor{Michael G. Jones}
\email{jonesmg@arizona.edu}

\author[0000-0002-5434-4904]{Michael G. Jones}
\affiliation{Steward Observatory, University of Arizona, 933 North Cherry Avenue, Rm. N204, Tucson, AZ 85721-0065, USA}

\author[0000-0003-4102-380X]{David J. Sand}
\affiliation{Steward Observatory, University of Arizona, 933 North Cherry Avenue, Rm. N204, Tucson, AZ 85721-0065, USA}

\author[0000-0001-9649-4815]{Bur\c{c}in Mutlu-Pakdil}
\affil{Department of Physics and Astronomy, Dartmouth College, Hanover, NH 03755, USA}

\author[0000-0001-8245-779X]{Catherine E. Fielder}
\affiliation{Steward Observatory, University of Arizona, 933 North Cherry Avenue, Rm. N204, Tucson, AZ 85721-0065, USA}

\author[0000-0002-1763-4128]{Denija Crnojevi\'{c}}
\affil{University of Tampa, 401 West Kennedy Boulevard, Tampa, FL 33606, USA}

\author[0000-0001-8354-7279]{Paul Bennet}
\affiliation{Space Telescope Science Institute, 3700 San Martin Drive, Baltimore, MD 21218, USA}

\author[0000-0002-0956-7949]{Kristine Spekkens}
\affiliation{Department of Physics and Space Science, Royal Military College of Canada P.O. Box 17000, Station Forces Kingston, ON K7K 7B4, Canada}
\affiliation{Department of Physics, Engineering Physics and Astronomy, Queen’s University, Kingston, ON K7L 3N6, Canada}

\author[0000-0001-7618-8212]{Richard Donnerstein}
\affiliation{Steward Observatory, University of Arizona, 933 North Cherry Avenue, Rm. N204, Tucson, AZ 85721-0065, USA}

\author[0000-0001-9775-9029]{Amandine Doliva-Dolinsky}
\affil{University of Tampa, 401 West Kennedy Boulevard, Tampa, FL 33606, USA}
\affil{Department of Physics and Astronomy, Dartmouth College, Hanover, NH 03755, USA}

\author[0000-0001-8855-3635]{Ananthan Karunakaran}
\affiliation{Department of Astronomy \& Astrophysics, University of Toronto, Toronto, ON M5S 3H4, Canada}
\affiliation{Dunlap Institute for Astronomy and Astrophysics, University of Toronto, Toronto ON, M5S 3H4, Canada}

\author[0000-0002-1468-9668]{Jay Strader}
\affil{Center for Data Intensive and Time Domain Astronomy, Department of Physics and Astronomy, Michigan State University, East Lansing, MI 48824, USA}

\author[0000-0002-5177-727X]{Dennis Zaritsky}
\affiliation{Steward Observatory, University of Arizona, 933 North Cherry Avenue, Rm. N204, Tucson, AZ 85721-0065, USA}



\begin{abstract}

We report the discovery of Corvus~A, a low-mass, gas-rich galaxy at a distance of approximately 3.5~Mpc, identified in DR10 of the Dark Energy Camera Legacy Imaging Survey during the initial phase of our ongoing SEmi-Automated Machine LEarning Search for Semi-resolved galaxies (SEAMLESS). Jansky Very Large Array observations of Corvus~A detect \hi \ line emission at a radial velocity of $523\pm2$~\kms. Magellan/Megacam imaging reveals an irregular and complex stellar population with both young and old stars. We detect UV emission in Neil Gehrels Swift observations, indicative of recent star formation. However, there are no signs of \hii \ regions in H$\alpha$ imaging from Steward Observatory's Kuiper telescope. Based on the Megacam color magnitude diagram we measure the distance to Corvus~A via the tip-of-the-red-giant-branch standard candle as $3.48\pm0.24$~Mpc. This makes Corvus~A remarkably isolated, with no known galaxy within $\sim$1~Mpc. Based on this distance, we estimate the \hi \ and stellar mass of Corvus~A to be $\log M_\mathrm{HI}/\mathrm{M_\odot} = 6.59$ and $\log M_\ast/\mathrm{M_\odot} = 6.0$. 
Although there are some signs of rotation, the \hi \ distribution of Corvus~A appears to be close to face-on, analogous to that of Leo~T, and we therefore do not attempt to infer a dynamical mass from its \hi \ line width. Higher resolution synthesis imaging is required to confirm this morphology and to draw robust conclusions from its gas kinematics.

\end{abstract}



\section{Introduction} \label{sec:intro}

Low-mass dwarf galaxies are the most numerous galaxies in the universe. They include the most dark matter-dominated galaxies and those most easily affected by environment. Dwarf galaxies have thus been the repeated focus of tensions (and resolutions) between observations and simulations of galaxy formation and cosmology, and are a key testing ground for our understanding of galaxy evolution \citep[see][for reviews]{Bullock+2017,Sales+2022}.

The lowest mass galaxies (e.g. $M_\ast \lesssim 10^6$~\Msol) are challenging to detect and until recently they could mostly only be identified within the Local Group (LG). Our understanding of how these galaxies form has therefore necessarily developed along with an appreciation for how the environment within the LG affects their present day properties \citep[e.g.][]{Weisz+2011,Spekkens+2014,Sawala+2016,Wetzel+2016,Grand+2017,Applebaum+2021,Putman+2021}. However, with recent advances in instrumentation, data processing, and search techniques it is now possible to identify these low-mass galaxies well beyond the LG and this necessity is beginning to fade \citep[e.g.][]{Chiboucas+2013,Abraham+2014,Crnojevic+2014,Crnojevic+2016,crnojevic19,Carlin+2016,Smercina+2017,Smercina+2018,Bennet+2019,Muller+2019,Carlin+2021,Carlsten+2022,Mutlu-Pakdil+2022,Mutlu-Pakdil+2024,McNanna+2023,Li+2024}. By leveraging these new capabilities it will soon be possible to construct a statistical sample of isolated extremely low mass galaxies, where the confounding effects of environment are minimized, which in turn will provide some of the cleanest and most robust tests of dark matter halo structure and cosmological processes (i.e. cosmic reionization).

The lowest mass galaxies in the LG have almost exclusively been identified with resolved star searches \citep[e.g.][]{Willman+2005,Belokurov+2006,Irwin+2007,McConnachie+2008,Koposov+2015,Drlica-Wagner+2015,Cerny+2021}, but similar mass galaxies outside of the LG can no longer be resolved into stars (due to their distance) in most wide-field, ground-based imaging surveys \citep[e.g.][]{Sand+2022}. However, within the distance range where existing surveys still have the requisite sensitivity to detect extremely low mass galaxies (e.g. $\log M_\ast/\mathrm{M_\odot} \lesssim 6$), these galaxies are not fully diffuse, but rather exist in a semi-resolved regime where their brightest stars appear as point sources on top of diffuse light from unresolved stars. In this regime neither the well-established resolved star searches nor standard integrated light searches perform well \citep{Mutlu-Pakdil+2021}. The advent of the Rubin Observatory's Legacy Survey of Space and Time (LSST) and the Nancy Grace Roman Space Telescope's High Latitude Wide Area Survey will push the resolved regime out further than ever before (in a wide-area survey), but there are still undoubtedly semi-resolved, extremely low mass galaxies (both quenched and star-forming) within existing surveys that could be identified with a well-suited search technique.

In a recent letter \citep{Jones+2023b} we used the combination of an integrated light search, a neural network classifier, and, finally, visual classification, to identify the Pavo dwarf galaxy, currently the lowest mass star-forming galaxy known in isolation. This search effort has uncovered several other strong candidates for nearby quenched and star-forming galaxies in this mass regime ($\log M_\ast/\mathrm{M_\odot} \lesssim 6$). Here we present another of our initial discoveries, a slightly more massive, blue, irregular galaxy in the constellation Corvus, which we will refer to as Corvus~A. In the following section we briefly discuss the discovery and follow-up observations of Corvus~A. In \S\ref{sec:props} we present our measurement of its distance, physical properties, and environment. In \S\ref{sec:discussion} we discuss the significance of these properties, focusing primarily on its \hi \ distribution and velocity field. Finally, we draw our conclusions in \S\ref{sec:conclusion}.

\section{Discovery and follow-up} \label{sec:search+follow-up}

Corvus~A was identified in the Dark Energy Spectroscopic Instrument (DESI) legacy imaging surveys \citep{Dey+2019} during the early development of our SEmi-Automated Machine LEarning Search for Semi-resolved galaxies (SEAMLESS). SEAMLESS has revealed numerous nearby, extremely low mass galaxies candidates in the legacy survey footprint, including the Pavo dwarf galaxy \citep{Jones+2023b}. Corvus~A was recovered from the Dark Energy Camera Legacy Survey (DECaLS) imaging in a new region of the survey footprint, included as part of DR10. A description of the SEAMLESS search technique is given in \citet{Jones+2023b}. Here we provide a brief overview. 

Our approach is built on that of \citet{Zaritsky+2019,Zaritsky+2023} which begins by masking high surface brightness emission from stars and bright galaxies and then filtering the images on a variety of scales in order to identify faint, extended sources. This returns millions of candidates over the entire survey footprint, which are whittled down first by cuts based on the \texttt{GALFIT} \citep{Peng+2002,Peng+2010} source parameter values and then by a convolutional neural network (CNN) classifier. The CNN used by SMUDGes was retrained for SEAMLESS specifically to identify nearby ($D<4$~Mpc) resolved and semi-resolved low-mass galaxies and to reject spurious detections from Galactic cirrus and distant background galaxy groups. This resulted in a sample of around 1000 candidates, which were then visually inspected by the team to identify the most compelling extremely low-mass galaxy candidates for follow-up. The visual inspection resulted in a final sample of 321 candidates, most of which are known nearby galaxies, and we are in the process of following up a few dozen new and high confidence candidates.

\subsection{Magellan Megacam imaging} \label{sec:magellan}

\begin{figure}
    \centering
    \includegraphics[width=\columnwidth]{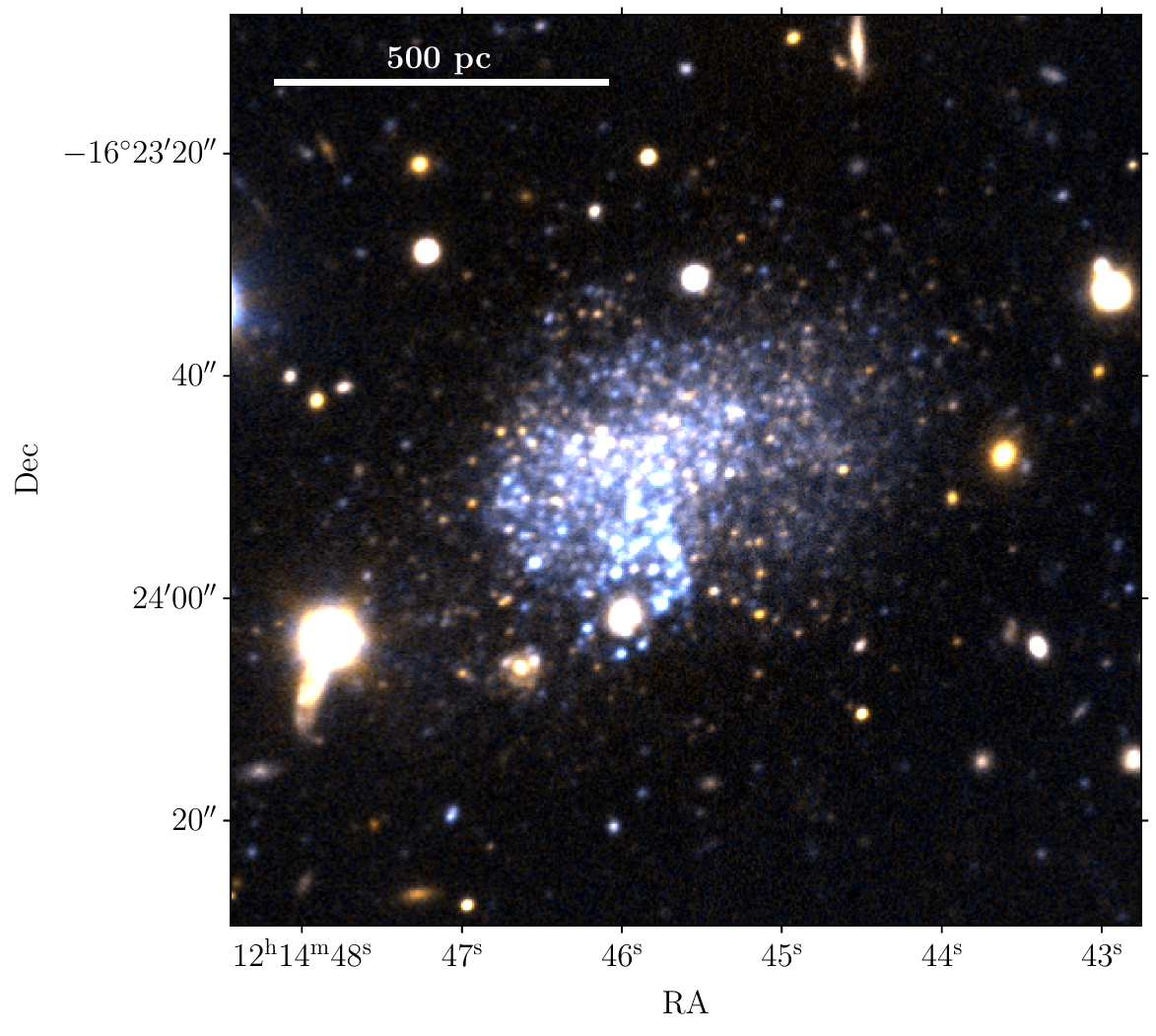}
    \caption{False color $g+r$ cutout from the Megacam image of Corvus~A. The blue region of recent star formation is clearly visible on the eastern side of the galaxy, but there is also an underlying redder population of stars.}
    \label{fig:Megacam}
\end{figure}

Deep follow-up imaging of Corvus~A was obtained with the Megacam imager \citep{megacam} on the Magellan Clay telescope (Figure~\ref{fig:Megacam}). The field of view for Megacam is $\sim$24\arcmin$\times$24\arcmin, with a 0.16 arcsec/pixel binning. Observations were taken the night of January 20, 2023 (part of 2023A-ARIZONA-1; PI: C.~Fielder) in $g$ (6 $\times$ 300~s) and $r$ (6 $\times$ 300~s) bands in $\approx$0.7\arcsec \ seeing. We reduced the data using the Megacam pipeline developed at the Harvard-Smithsonian Center for Astrophysics by M. Conroy, J. Roll, and B. McLeod, including detrending the data, performing astrometry, and stacking the individual dithered frames.

\begin{figure*}
    \centering
    \includegraphics[width=\textwidth]{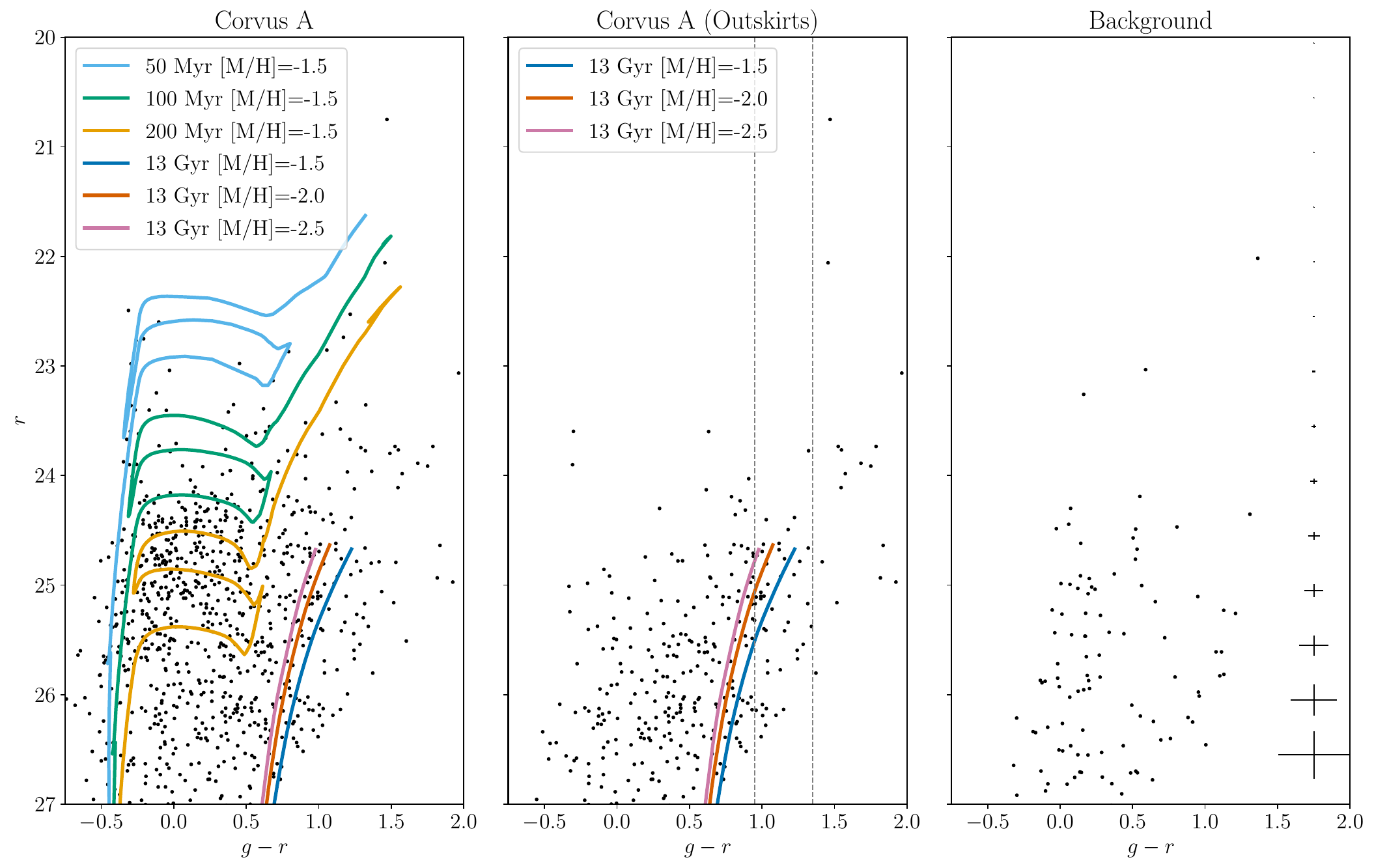}
    \caption{\textit{Left}: CMD of Corvus~A from the Megacam imaging. Stars were selected within the large circular aperture (Appendix~\ref{sec:aperture}) and as described in \S\ref{sec:magellan}. PARSEC isochrones \citep{Bressan+2012} shown for stellar populations of age 50, 100, and 200~Myr, and Dartmouth isochrones \citep{Dotter+2008} are plotted for the RGB of an ancient population with three different metallicities. These are all shifted in magnitude by a distance modulus of 27.71 (i.e. 3.48~Mpc; \S\ref{sec:dist}). \textit{Center}: CMD of stars in the outskirts of Corvus~A, selected as described in \S\ref{sec:dist}. RGB isochrones are plotted as in the left panel. The grey dashed vertical lines indicate the region used to measure the TRGB. \textit{Right}: CMD of a region of the Megacam image far from Corvus~A. Point sources were selected in an equal area region to that of the left panel, which is approximately 1.4 times larger than the area used for the central panel. The typical photometric uncertainties (for all panels) are plotted along the rightmost edge of the figure.}
    \label{fig:CMD}
\end{figure*}

We perform point-spread function (PSF) fitting photometry on the final image stacks, using the DAOPHOT and ALLFRAME software suite \citep{Stetson87,Stetson94}, following the same methodology described in \citet{Mutlu-Pakdil+2021}. We remove objects that are not point sources by culling our catalogs of outliers in $\chi^2$ versus magnitude, magnitude error versus magnitude, and sharpness versus magnitude space. Instrumental magnitudes are then calibrated to the DECaLS DR10 catalog. We correct for Galactic extinction on a star-by-star basis using the \citet{Schlegel1998} reddening maps with the coefficients from \citet{Schlafly2011}, which were evaluated according to the \citet{Fitzpatrick1999} reddening law with normalization $N = 0.78$. Specifically, we adopted the coefficients of 3.237 and 2.176 for $g$ and $r$ to convert E(B-V) to $A_g$ and $A_r$, respectively. The resulting color magnitude diagram (CMD) of Corvus~A is shown in Figure~\ref{fig:CMD}, which is discussed further in \S\ref{sec:dist} \& \S\ref{sec:stellar_pop}.

\begin{figure*}
    \centering
    \includegraphics[width=0.44\textwidth]{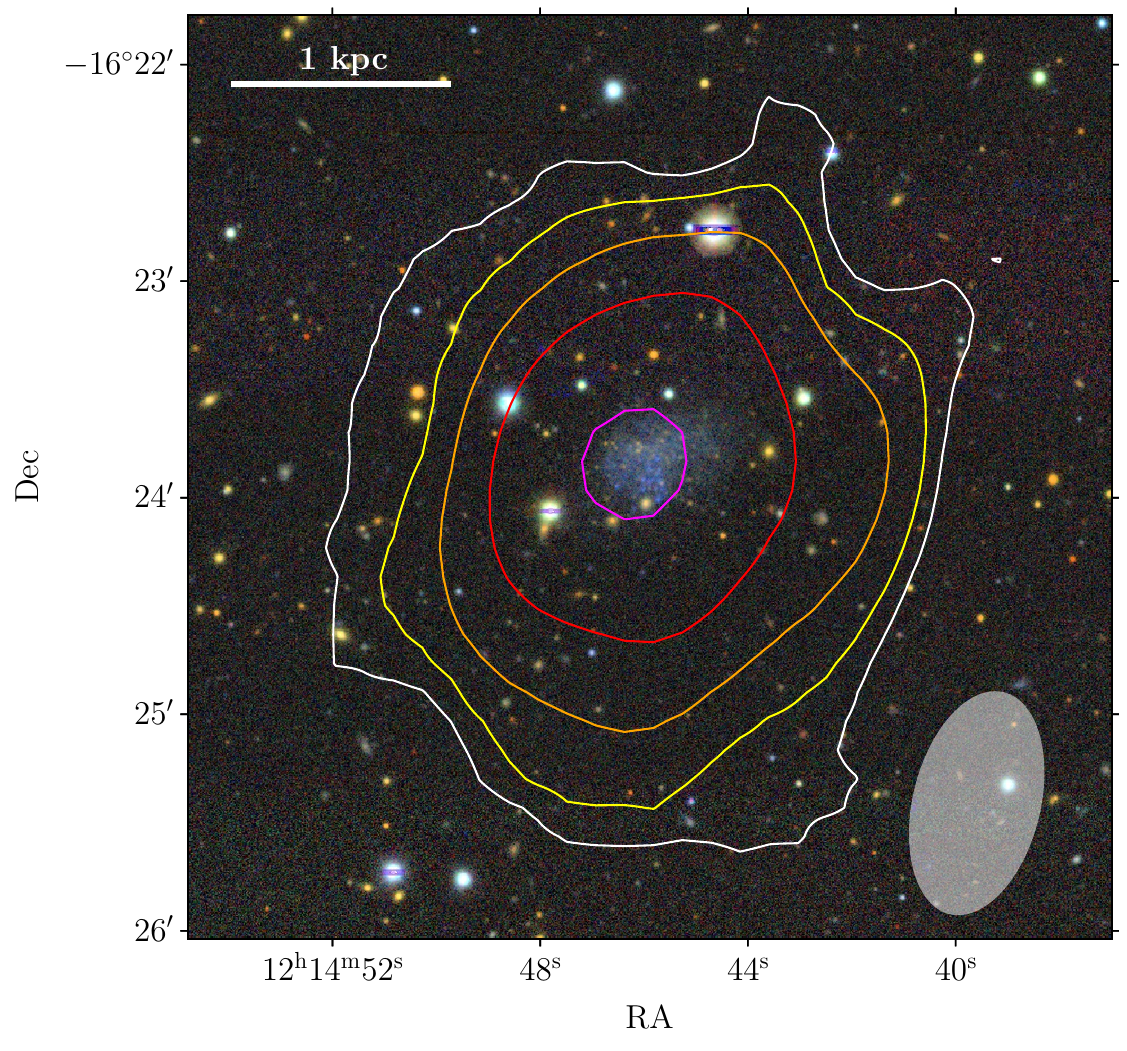}
    \includegraphics[width=0.55\textwidth]{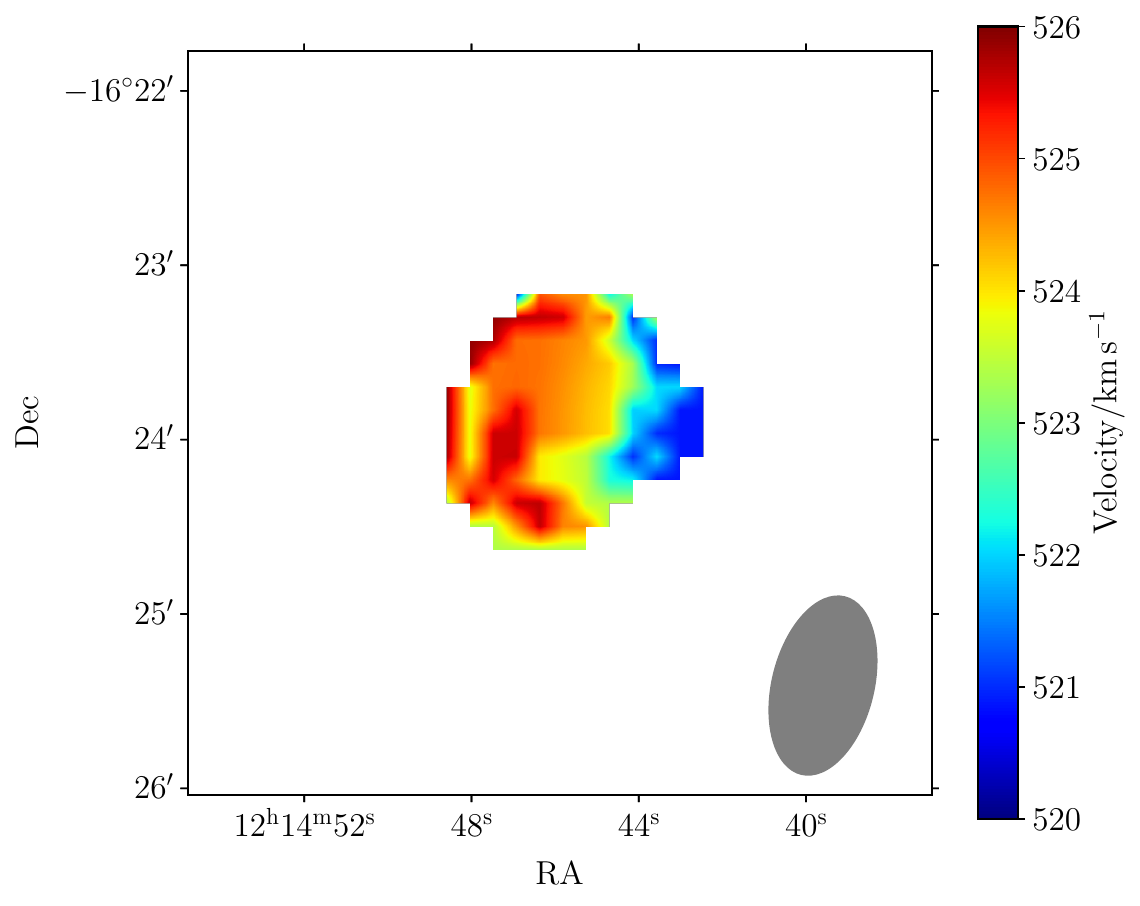}
    \caption{\textit{Left}: VLA \hi \ moment zero contours overlaid on a DECaLS $griz$ image. The contours begin at $1.2 \times 10^{19} \; \mathrm{cm^{-2}}$ (0.095~\Msol~pc$^{-2}$) over 10~\kms, and each subsequent contour is double the previous. The grey ellipse in the lower right shows the synthesized beam. \textit{Right}: Moment one map of \hi \ emission in Corvus~A. The FoV is the same as in the left panel.}
    \label{fig:Corvus_VLA}
\end{figure*}

\subsection{VLA observations} \label{sec:vla}

\begin{figure}
    \centering
    \includegraphics[width=\columnwidth]{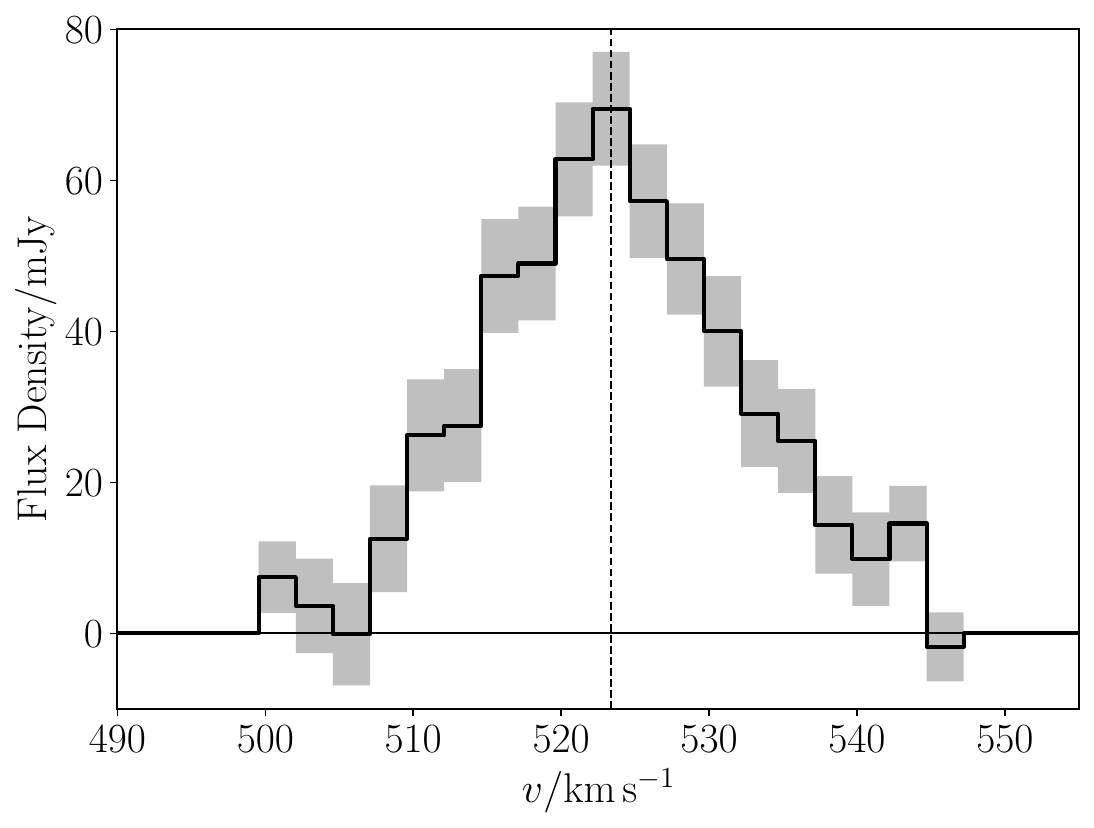}
    \caption{\hi \ line profile of Corvus~A from the VLA observation. Only emission within the \texttt{SoFiA} mask is included. The grey band shows 1$\sigma$ uncertainty in each channel based on the rms noise and the number of pixels included in the mask. The vertical dotted line marks the line center (flux-weighted centroid).}
    \label{fig:HI_spec}
\end{figure}

Shortly after Corvus~A was discovered we identified a tentative \hi \ signal in the \hi \ Parkes All Sky Survey \citep[HIPASS;][]{Barnes+2001}. 
We obtained VLA D-array observations (23B-188; PI: M.~Jones) to verify this signal and to confirm that it is localized to Corvus~A, as the Parkes telescope has $\sim$13\arcmin \ resolution L-band. In total the VLA was on-source for 1.25~h and data were recorded with a native spectral resolution of 3.9~kHz over a total bandwidth of 4~MHz ($\sim$800~\kms), centered on the \hi \ line frequency (redshifted by 500~\kms). Data were reduced following standard tasks in the Common Astronomy Software Applications \citep[\texttt{CASA};][]{CASA} package within the pipeline\footnote{\url{https://github.com/AMIGA-IAA/hcg_hi_pipeline}} of \citet{Jones+2023}. During the final imaging we regridded the spectral channels to a resolution of 2.5~\kms. The final synthesized beam size is 76.0\arcsec$\times$43.5\arcsec \ and the rms noise is 2.3~mJy/beam in 2.5~\kms \ channels (both for robust=0.5 weighting). The Source Finding in Astronomy tool \citep[\texttt{SoFiA};][]{Serra+2015} was used for source masking, with spatial smoothing over approximately one beam size and a combination of no spectral smoothing and smoothing over 20~km/s, with a 4$\sigma$ threshold for inclusion in the mask. The moment zero map of \hi \ emission is shown in Figure~\ref{fig:Corvus_VLA} (left) overlaid on a DECaLS $griz$ image, and the line profile for all the emission included in the \texttt{SoFiA} mask is shown in Figure~\ref{fig:HI_spec}. We also produced a moment one map (Figure~\ref{fig:Corvus_VLA}, right) by taking the \texttt{SoFiA} mask and further masking any pixel with an intensity below six times the rms noise. This significantly helps to reduce noise in the moment map, but has the effect of shrinking the area sampled. Thus the region with data in Figure~\ref{fig:Corvus_VLA}, (right) is significantly smaller than in the left panel, even though they show the same FoV. From this map we can see a clear velocity gradient across Corvus~A, indicating that it is likely rotating, with the eastern side receding and the western side approaching us. However, the resolution is too poor to permit kinematic modeling.

\subsection{Kuiper telescope H$\alpha$ imaging}

\begin{figure*}
    \centering
    \includegraphics[width=\textwidth]{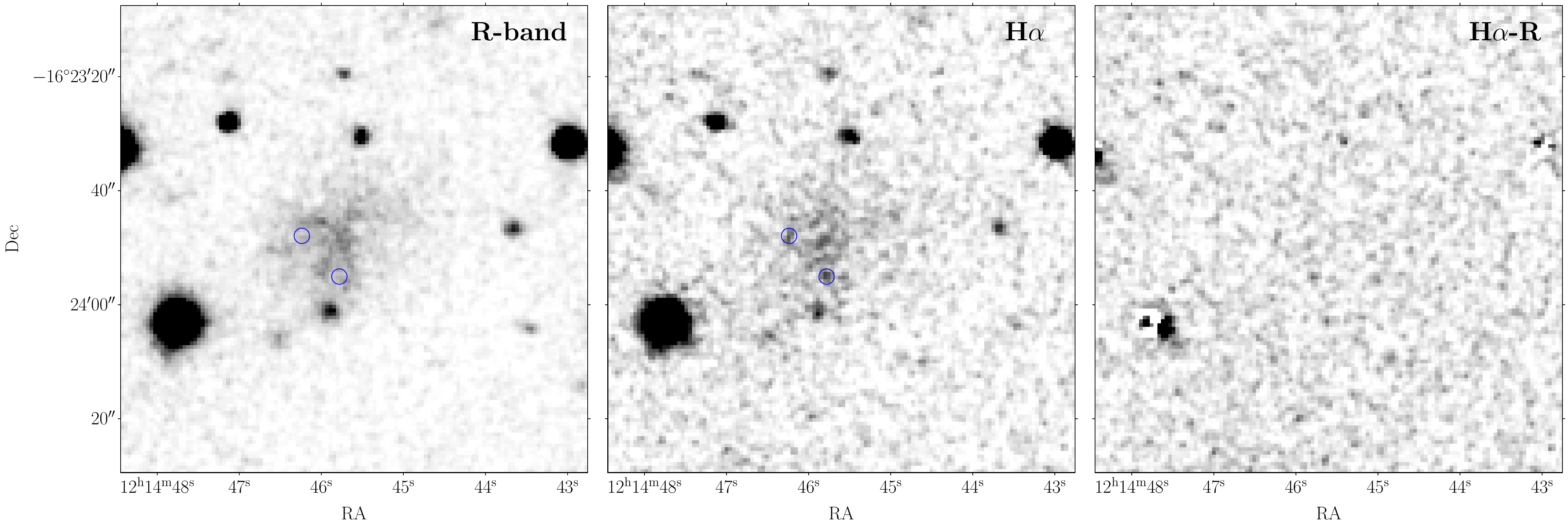}
    \caption{\textit{Left}: Harris R-band image of Corvus~A from Kuiper/Mont4K covering the same FoV as Figure~\ref{fig:Megacam}. \textit{Center}: Corresponding narrow band H$\alpha$ image. \textit{Right}: H$\alpha$ image with the R-band continuum subtracted. In the center (and left) panel two apparent faint clumps in the H$\alpha$ image, which do not seem to have direct counterparts in the R-band image, are highlighted with blue circles. However, the brighter of these (more southern) showed no emission lines when followed up with a SOAR spectrum.}
    \label{fig:Kuiper_Ha}
\end{figure*}

Corvus~A was observed in Harris R-band and narrow band H$\alpha$ with the Kuiper 61-inch telescope at Mt. Bigelow Observatory in April 2023. Five 100~s R-band exposures and $10 \times 250$~s H$\alpha$ exposures were taken with the Mont4K imager in good conditions. The resulting co-added images and the R-band-subtracted H$\alpha$ image are shown in Figure~\ref{fig:Kuiper_Ha}. Almost all of the emission seen in the H$\alpha$ image appears to be consistent with continuum and is removed by the R-band subtraction. Thus, Corvus~A does not appear to have any \hii \ regions or ongoing star formation.

Despite there being no clear \hii \ regions, there are two faint clumps in the H$\alpha$ image that do not have clear counterparts in the R-band image. In May 2024 we obtained an optical spectrum of the brighter of these clumps with the Southern Astrophysical Research (SOAR) Telescope, but there were no emission lines, confirming the lack of H$\alpha$ emission.

\subsection{Swift UV observations}

\begin{figure}
    \centering
    \includegraphics[width=\columnwidth]{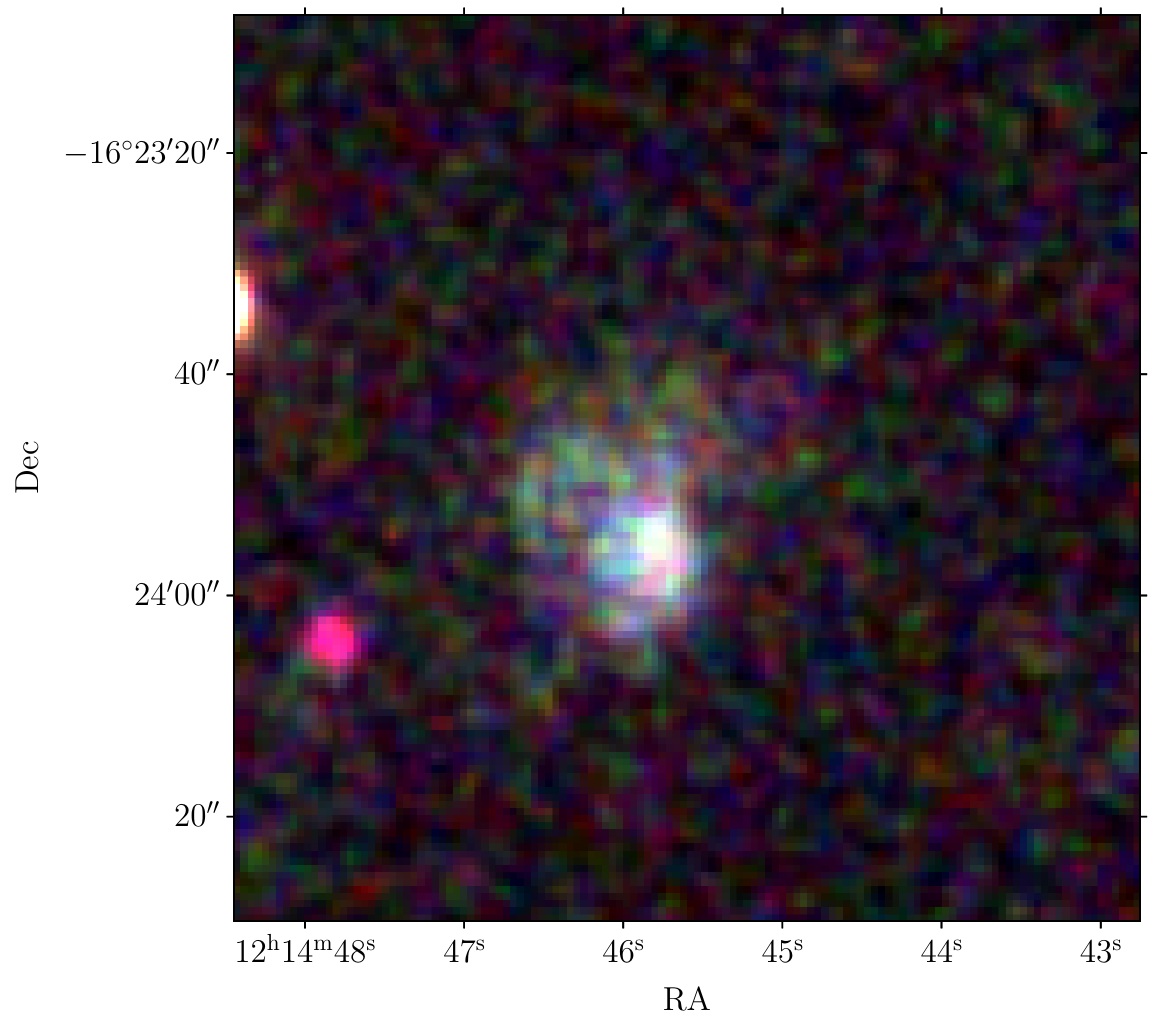}
    \caption{Swift UVW1, UVM1, UVW2 composite color image of Corvus~A showing the same FoV as Figure~\ref{fig:Megacam}.}
    \label{fig:Swift}
\end{figure}

Although the lack of H$\alpha$ implies that Corvus~A does not have ongoing star formation, it clearly contains young, blue stars. In order to estimate Corvus~A's star formation rate in the recent past (i.e. 100-200 Myr) we use UV data. As Corvus~A is not in the Galaxy Evolution Explorer (GALEX) all-sky imaging survey footprint, we obtained UV imaging with the Ultra-Violet Optical Telescope \citep[UVOT;][]{Roming+2005} on the NASA Neil Gehrels Swift observatory \citep{Gehrels+2004} in all three UV bands (UVW1, UVM2, UVW2). Corvus~A was observed in a single sequence on March 31, 2023 with an exposure time of $\sim$300~s in each band. Corvus~A is clearly detected as excess UV emission in all three bands (see Figure \ref{fig:Swift}).

\section{Properties of Corvus~A} \label{sec:props}

\begin{table}[]
    \centering
    \caption{Properties of Corvus~A}
    \begin{tabular}{lc}
    \hline\hline
    Parameter       &  Value\\ \hline
    RA              &  12:14:45.6 \\
    Dec.            & $-$16:23:49 \\
    $m_g$/mag           & $16.8\pm0.1$ \\
    $m_r$/mag           & $16.5\pm0.1$ \\
    $S_{21}$/Jy~\kms      & $1.36\pm0.05$\\
    $v_\mathrm{HI}$/\kms  & $523\pm2$ \\
    Dist./Mpc       & $3.48 \pm 0.24$ \\
    $M_V$/mag           & $-11.2\pm0.2$ \\
    $r_h$           & 15.2\arcsec$\,\pm\,$0.5\arcsec \\
    $r_h$/pc        & $256 \pm 19$\\
    $a/b$      & $1.43 \pm 0.03$ \\
    $\epsilon$ & $0.30 \pm 0.01$ \\
    $\theta$        & $110^\circ \pm 10^\circ$ \\
    $\log (\mathrm{SFR_{UV}/M_\odot \, yr^{-1}})$ & $-3.25\pm0.07$ \\
    $\log M_\ast$/\Msol  & $6.0\pm0.1$ \\
    $\log M_\mathrm{HI}$/\Msol & $6.59\pm0.06$ \\
    \hline
    \end{tabular}
    \label{tab:props}
\end{table}

\subsection{Color and morphology} \label{sec:morph}

Corvus~A has a blue, clumpy, and irregular appearance reminiscent of Leo~P \citep{Rhode+2013} and galaxies in the Survey of \hi \ in Extremely Low-mass Dwarfs \citep[SHIELD;][]{Cannon+2011}. Even in the DECaLS imaging, Corvus~A's light distribution is markedly clumpy, indicating that it is almost resolved into stars, which in turns implies that it must be within a few Mpc. In the Megacam image (Figure~\ref{fig:Megacam}) we see that Corvus~A is almost entirely resolved into individual stars, though there are still regions with significant crowding. In particular, the south-east of the stellar body is dominated by bright, blue stars that likely are the result of Corvus~A's most recent star formation episode. However, as shown by the H$\alpha$ imaging, there are no \hii \ regions, indicating that this star formation must have occurred over $\sim$10~Myr ago. This region of Corvus~A is also where the UV emission (Figure~\ref{fig:Swift}) is concentrated, again highlighting this as recent star formation. The western side of Corvus~A is undetected in UV, suggesting that the star formation over the past $\sim$200~Myr was confined to the south-east of the galaxy. Furthermore, in the \hi \ map (Figure~\ref{fig:Corvus_VLA}, left) we see that the neutral gas content of Corvus~A is also centered on the eastern side.

\subsection{Distance} \label{sec:dist}

To measure the distance to Corvus~A we use the discontinuity at the tip-of-the-red-giant-branch (TRGB) in the CMD (Figure~\ref{fig:CMD}) as a standard candle \citep[e.g.,][]{dacosta90, lee93, makarov06}. Corvus~A is quite irregular and there is considerable crowding within its region of recent star formation. We therefore manually construct an aperture (Appendix~\ref{sec:aperture}) to produce a clean CMD of Corvus~A (Figure~\ref{fig:CMD}, center panel) from which to measure the TRGB. 

To locate the position of the TRGB we first removed all stars bluer than $g-r=0.95$ to eliminate any remaining young stars, and a second cut of $g-r<1.35$ was applied to help remove foreground stars. The observed luminosity function of the red giant branch (RGB) stars was then measured from the CMD and fit with a model luminosity function after applying photometric uncertainties, bias, and completeness determined with artificial star tests \citep{crnojevic19}. Using a nonlinear least squares fit between this model and the observed luminosity function we calculated the apparent magnitude of the TRGB as $r_\mathrm{TRGB} = 24.70\pm0.12$, which results in a distance modulus of $27.71 \pm 0.15$, based on an $r$-band TRGB calibration of $M_{r}^\mathrm{TRGB} = -3.01 \pm 0.01$ \citep{sand14}. This in turn equates to a distance of $3.48 \pm 0.24$~Mpc.

We note that this measurement has relatively large uncertainty as the RGB is not very well populated in the regions of the galaxy where we can cleanly select RGB stars. However, the location of the TRGB that we have measured appears to be consistent with the initial CMD from recent Hubble Space Telescope (HST) observations of Corvus~A (Mutlu-Pakdil et al. in prep.).

\subsection{Gas content} \label{sec:gas}

In Figure~\ref{fig:Corvus_VLA} we see that the peak of the \hi \ distribution of Corvus~A is centered on the site of the most recent star formation, on the eastern side of the stellar body. Although the beam size of the VLA observation is larger than the visible extent of the stellar body, for high S/N detections the centroiding accuracy is considerably better than the beam resolution, and the \hi \ distribution clearly peaks to the east of the optical center. The \hi \ distribution is also marginally resolved, extending over approximately five beam widths in the east-west direction, but only two in the north-south direction (as the beam is elongated north-south). At this resolution we can only discern a mild velocity gradient in the moment one map (Figure~\ref{fig:Corvus_VLA}, right), with the eastern side of Corvus~A being the receding side and the western side the approaching side. However, the line profile has a triangular shape, perhaps indicating that rotation is not the dominant component broadening the line (\S\ref{sec:faceon}). The central velocity of this profile is 523~\kms, and the W50 and W20 values are 17.9 and 35.0~\kms, respectively. Integrating the full profile we obtain a total \hi \ flux of 1.36~Jy~\kms, which equates to an \hi \ mass of $\log M_\mathrm{HI}/\mathrm{M_\odot} = 6.59$. Thus, Corvus~A is roughly five times more massive (in \hi) than Leo~P and represents an intermediate object between Leo~P and Pavo, and the low-mass galaxies in the SHIELD sample.

The peak \hi \ column density of Corvus~A in the VLA moment zero map is $2.1 \times 10^{20} \; \mathrm{cm^{-2}}$. At the level of $1 \times 10^{20} \; \mathrm{cm^{-2}}$ the \hi \ diameter of Corvus~A is approximately 79\arcsec \ (in the east-west direction). Correcting for beam smearing (based on the minor axis of the beam only) we obtain a value of 67\arcsec. Based on the \hi \ size--mass relation \citep{Wang+2016} a galaxy of $\log M_\mathrm{HI}/\mathrm{M_\odot} = 6.59$ at 3.5~Mpc is expected to have an \hi \ diameter of 65\arcsec, in close agreement with our approximate measurement.

\subsection{Photometry and structural parameters}\label{sec:struct}

\begin{figure}
    \centering
    \includegraphics[width=\columnwidth]{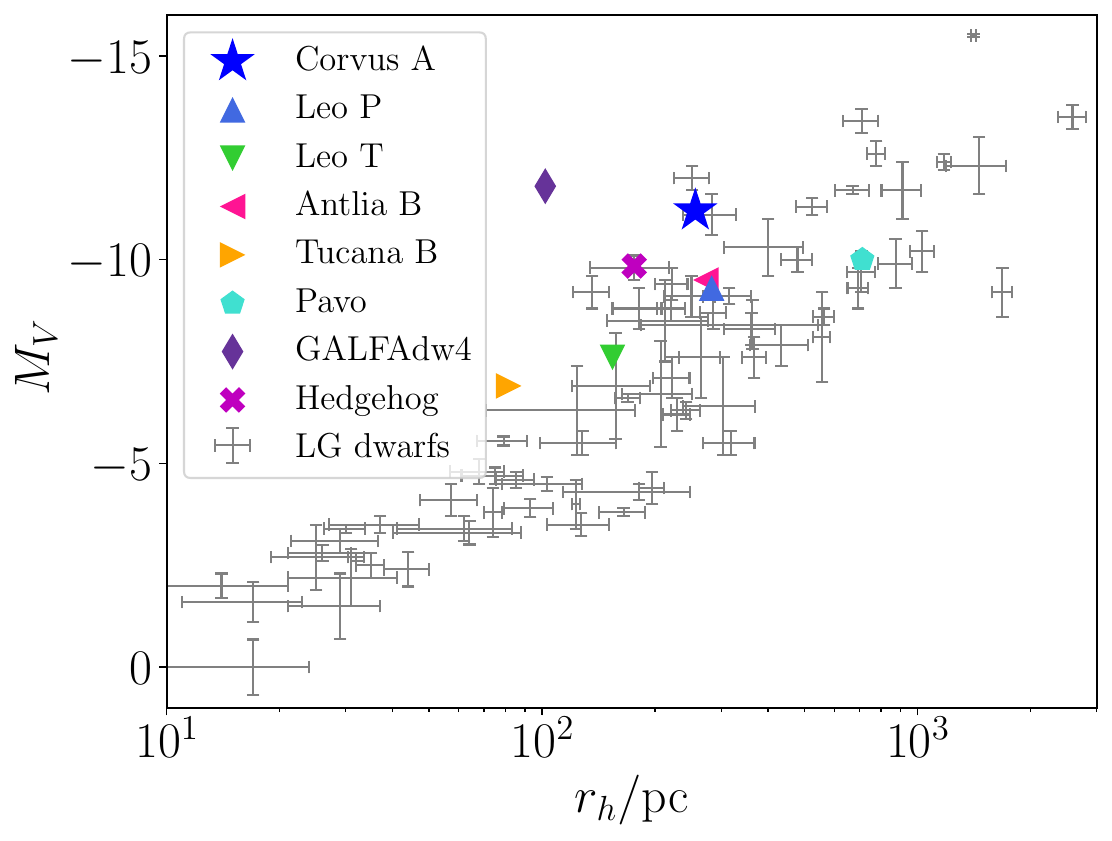}
    \caption{Size--luminosity relation of LG dwarfs with Corvus~A and other notable Local Volume dwarfs highlighted \citep{Irwin+2007,Giovanelli+2013,McQuinn+2015b,Sand+2015b,Sand+2022,Bennet+2022,Jones+2023b,Li+2024}.}
    \label{fig:sizemass}
\end{figure}

As Corvus~A has an irregular morphology with an off-center star-forming region that is severely crowded in the Megacam images, it poses significant difficulties for standard approaches for determining structural parameters. We attempted to construct a RGB star map to trace just the old stellar population, but no RGB stars in the direction of the star-forming region could be identified due to the crowding. Fitting in integrated light is similarly problematic as Corvus~A is both partially resolved into stars and its brightest region is off-center and irregularly shaped. The process to determine its structural parameters and photometry were thus necessarily semi-manual, as described below.

We began by determining the approximate center, axial ratio, and position angle of Corvus~A (Table~\ref{tab:props}) by manually constructing an ellipse that roughly aligns with isophotes near the visible edge of the galaxy (beyond the star-forming region) in the $r$-band image. We then masked sources in the vicinity of Corvus~A that appear to be background galaxies or foreground stars and measured the radial profile in concentric elliptical apertures. The integrated magnitudes were measured at the point that the profile reached the sky level, and the half-light radius was measured (in $r$-band) as the semi-major axis of the ellipse containing half of this integrated light. The uncertainties were estimated as half the difference between the half-light radius in $r$ and $g$. The E(B-V) reddening in the direction of Corvus~A is 0.0428 \citep{Schlafly2011}, which we have corrected for in the magnitudes in Table~\ref{tab:props}.

In Figure~\ref{fig:sizemass} we show Corvus~A on the size--luminosity relation in comparison to LG dwarfs and several notable dwarfs within the Local Volume. We see that like Leo~P, Antlia~B, and Tucana~B, Corvus~A is quite compact given its luminosity, but still within the scatter of the relation. This trend (of which Pavo is an exception) is likely the result of observational bias as more compact dwarfs (at a given luminosity) are more readily identifiable. An alternative explanation might be that the effective radii of these star-forming dwarfs are biased low (relative to quiescent dwarfs in the LG) due to the impact of clumpy recent star formation on their light distributions.

\subsection{Star formation rate and stellar mass estimates}

Although Corvus~A was detected in all three Swift/UVOT bands, we use only UVM2 to determine the star formation rate (SFR) as this is most similar to GALEX NUV \citep{Hoversten09}. The UV flux was measured within an elliptical aperture with a semi-major axis equal to two half-light radii (Table~\ref{tab:props}). This resulted in a UV magnitude of $m_\mathrm{UV} = 17.3 \pm 0.1$. This was then converted to a SFR following the relations of \citet{IglesiasParamo2006} for NUV, giving $\log \mathrm{SFR/M_\odot~yr^{-1}} = -3.25 \pm 0.07$ (including the distance uncertainty).

To estimate the total stellar mass of Corvus~A we use its $g-r$ color and magnitudes in the mass-to-light ratio relations of \citet{Zibetti+2009} and \citet{Into+2013}. These give stellar mass estimates of $\log M_\ast/\mathrm{M_\odot} = 5.89$ and 6.04, respectively. We take the log mean of these values as Corvus~A's stellar mass and their difference as an approximate indication of the uncertainty, i.e. $\log M_\ast/\mathrm{M_\odot} = 6.0 \pm 0.1$. As with its \hi \ mass this is a few times more massive than Leo~P (or Pavo). As the young stars are dominating the light from Corvus~A, it is quite likely that these mass-to-light ratio relations have under-counted the contribution from Corvus~A's old stellar population and that a more in-depth characterization of its stellar population would result in a slightly higher stellar mass estimate.

\subsection{Environment} \label{sec:env}

\begin{figure*}
    \centering
    \includegraphics[width=\columnwidth]{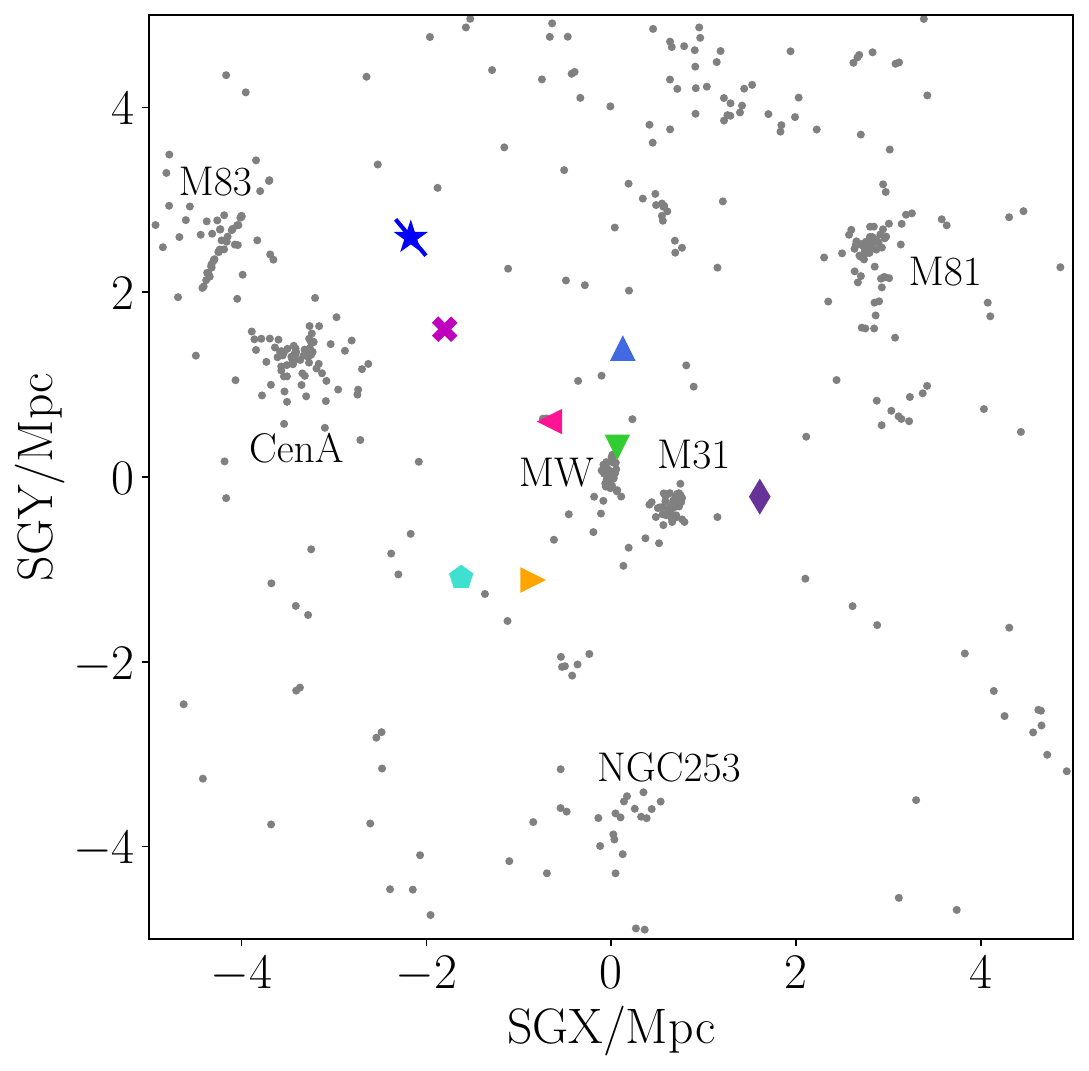}
    \includegraphics[width=\columnwidth]{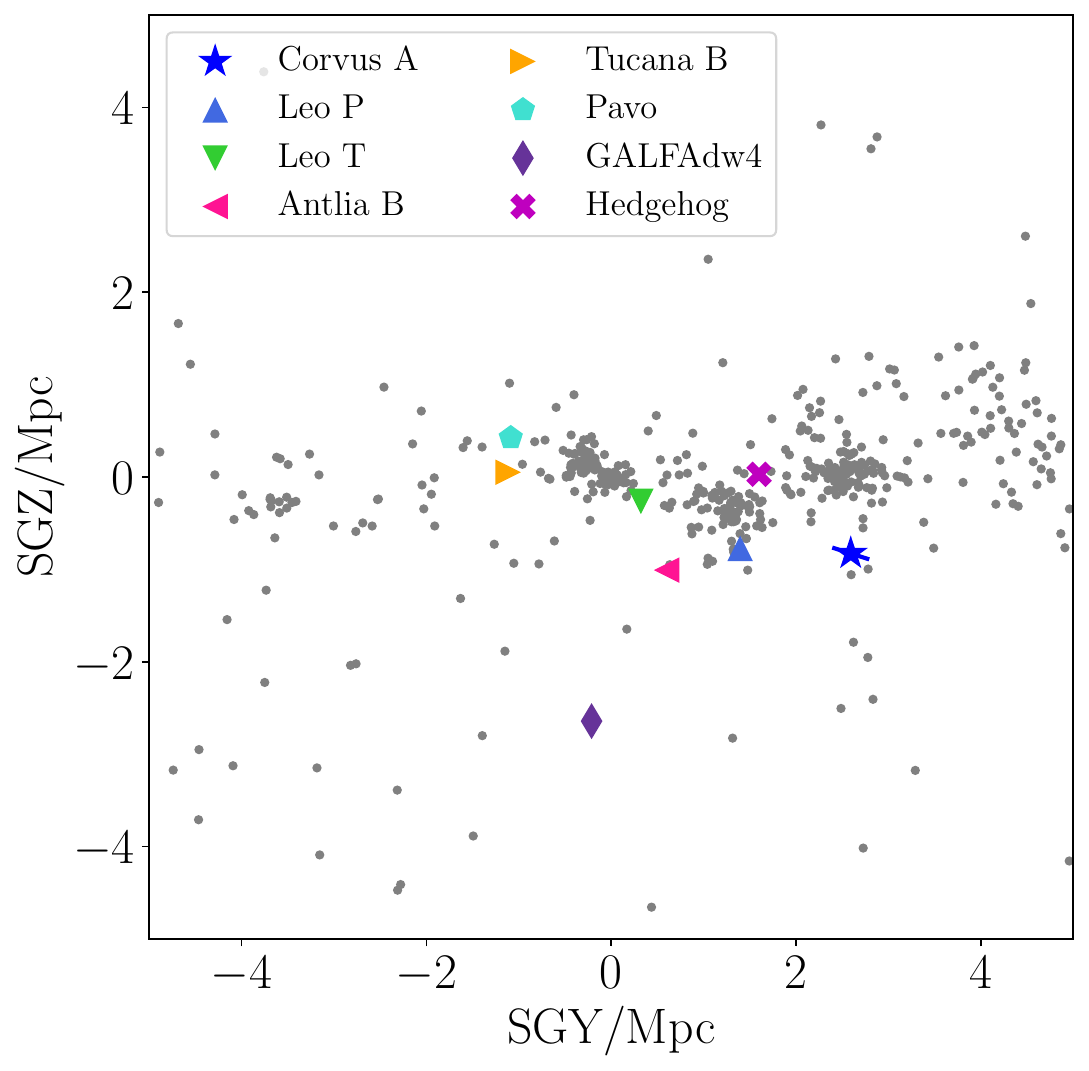}
    \caption{\textit{Left}: Supergalactic XY projection of Local Volume galaxies from the \citet{Karachentsev+2019} catalog. Corvus~A is shown by the blue star and errorbar. Several other notable low-mass galaxies are highlighted. \textit{Right}: Supergalactic YZ projection, otherwise as in the left panel.}
    \label{fig:env}
\end{figure*}

Figure~\ref{fig:env} shows the supergalactic XY and YZ projections of Local Volume galaxies \citep[from][]{Karachentsev+2019} within a $10\times10\times10$~Mpc cube centered on the Milky Way. Corvus~A is in a direction not associated with any major nearby structure. The nearest (in projection) significant structures are the Cen~A and M83 groups. Corvus~A is also at a similar recessional velocity to these groups ($v_\mathrm{CenA} = 547$~\kms \ and $v_\mathrm{M83} = 513$~\kms). However, they are both over 2~Mpc away and Corvus~A is unlikely to be associated with either. In three dimensions Corvus~A's nearest neighbor is dw1240-1140 at a separation of 930~kpc. This dwarf has a surface brightness fluctuation distance estimate of 4.25~Mpc \citep{Carlsten+2022}, however, that paper itself argues that this distance may be erroneous and dw1240-1140 may in fact be a satellite of M104 at $\sim$10~Mpc. Corvus~A's second nearest known neighbor is the blue, irregular dwarf galaxy PGC~1059300 at a 3D separation of 1~Mpc.

Corvus~A is thus considerably more isolated than either Pavo or Leo~P, both of which are regarded as isolated and pristine objects that have likely never been within the halo of a more massive system. However, unlike GALFAdw4 \citep{Bennet+2022}, Corvus~A still seems to be associated with the local sheet.

\section{Discussion} \label{sec:discussion}

\subsection{Stellar population} \label{sec:stellar_pop}

The Magellan/Megacam image of Corvus~A is extremely crowded in the region dominated by young stars, so it is not possible to perform a detailed assessment of its stellar population. However, a few basic points are still clear. Despite the crowding, the CMD contains numerous young stars, most likely blue and red He burning stars older than about 100~Myr, as is evident from the PAdova and TRieste Stellar Evolution Code \citep[PARSEC;][]{Bressan+2012} isochrones overlaid on Figure~\ref{fig:CMD} (left). The lack of H$\alpha$ emission suggests that there are no very young stars (e.g. $<10$~Myr) and this also seems to be consistent with the CMD. 

As in our discussion of Pavo \citep{Jones+2023b}, a bottom-heavy stellar initial mass function is also a possibility. Corvus~A could still be forming stars, but given the low SFR estimated from the UV emission there might still be no stars massive enough to form an \hii \ region. However, it is more likely that the star formation in Corvus~A is simply bursty and that, like Pavo, it could be entering a low point in its star formation history (SFH). It should be noted that while Pavo appears to have a meager \hi \ reservoir that was undetectable in archival wide-field data \citep{Jones+2023b}, Corvus~A is both gas-rich and several times higher stellar mass. The HST imaging of Corvus~A (Mutlu-Pakdil et al. in prep.) will permit a detailed accounting of the SFH over the past several hundred Myr, analogous to the SFHs of Local Volume dwarfs \citep[e.g.][]{Weisz+2011}.

Assuming that the lack of H$\alpha$ emission indicates that there is no current star formation in Corvus~A, it is worth considering whether this means that we are witnessing it at a particular time during its life cycle when it has young stars and UV emission but no ongoing star formation, while still containing a rich gas reservoir. The NUV is sensitive to star formation that occurred within the past $\sim$200~Myr \citep[e.g.][]{Lee+2009}, while H$\alpha$ emission only traces star formation within the past 10~Myr. Thus, the combination of NUV emission with no H$\alpha$ emission only places a weak timing constraint. 

The presence of a rich \hi \ reservoir while star formation has paused places a further timing constraint, however, this too is likely to be quite weak. The simulations of \citet{Rey+2022} suggest that galaxies such as Leo~P and Pavo may (temporarily) remove much of their \hi \ gas through stellar feedback after episodes of star formation. However, Corvus~A is considerably more massive and its \hi \ reservoir is unlikely to be rendered undetectable even after an episode of star formation. For comparison we note that 2/12 of the SHIELD galaxies (all of which are detected in \hi) presented in \citet{Teich+2016} have UV emission but no H$\alpha$, and these are several times more massive again. Thus, at masses $\log M_\ast/\mathrm{M_\odot} \lesssim 7$, this does not appear to be a particularly uncommon configuration and likely does not suggest that we are witnessing Corvus~A at a special time.

\subsection{A low-inclination, turbulence-broadened \hi \ distribution} \label{sec:faceon}

As there are only a handful of galaxies known with \hi \ masses similar to that of Corvus~A and we do see some indication of rotation in the moment one map (Figure~\ref{fig:Corvus_VLA}, right),  we considered attempting to place it on the baryonic Tully-Fisher relation (BTFR). However, this attempt highlighted a number of issues worth discussing briefly.

The \hi \ distribution in Corvus~A (Figure~\ref{fig:Corvus_VLA}, left) appears quite circular (even after considering the beam elongation). Assuming that the intrinsic shape of this gas distribution is roughly a circular disk, this suggests that its inclination is low ($\lesssim 30^\circ$).  
The axial ratio of the stellar body of Corvus~A (Table~\ref{tab:props}) implies a larger inclination. However, this may not be representative of the inclination of the gas disk for two reasons. First, in low-mass, star-forming galaxies the light of the stellar body can be dominated by the locations of clumpy stochastic star formation, which tends to result in a systematic overestimate of the disk inclination \citep{Read+2016}. Second, it is possible that the gas distribution and the stellar body do not have similar geometry. In particular, the major axis of the stellar body appears to be almost perpendicular to the major axis of the gas distribution.

In addition, the triangular shape of the global \hi \ line profile (Figure~\ref{fig:HI_spec}) of Corvus~A is unusual for an inclined rotating disk but characteristic of turbulent broadening by a two-component neutral medium \citep[][]{young96,young97,young03}. Accounting for instrumental broadening \citep[e.g.][]{Springob+2005}, the FWHM of Corvus~A's profile is only $\sim$16~\kms. A turbulent gas with a standard deviation of $ \sigma_\mathrm{turb} \sim 7-10$~\kms , characteristic of the $T\sim6000$~K warm neutral medium that dominates the \hi \ reservoirs of gas-rich low-mass galaxies \citep[e.g.][]{warren12,bc14,Adams+2018}, can therefore completely account for the width of Corvus~A's line profile, with rotational broadening making only a minor contribution.

The above discussion implies that Corvus~A's \hi \ distribution, if arranged in a disk, is likely close to face-on, in which case it is not possible to estimate a reliable rotation velocity with the currently available data. This is analogous to the situation for Leo~T \citep{Adams+2018}, which also has only weak signs of rotation in its moment one map, a roughly circular \hi \ moment zero map, and a turbulence-broadened \hi \ profile. With higher spectral and spatial resolution data (e.g. from MeerKAT or additional VLA configurations) it should be possible both to separate the broadening from the warm and cold neutral media, and to distinguish the contributions from turbulence and rotation, at which point it may be possible to fit a kinematic model and determine a rotation velocity if the inclination is not too small.

\section{Conclusions} \label{sec:conclusion}

Corvus~A is a newly discovered low-mass galaxy identified during the initial phase of the SEAMLESS project. The stellar body of Corvus~A has an irregular structure and its light is dominated by a region of young blue stars on its eastern side. Follow-up observations with the VLA, Magellan, Swift, and the Kuiper telescope have revealed that Corvus~A is gas-rich and although it contains many young stars ($<$200~Myr), it does not host any \hii \ regions. We have measured the distance to Corvus~A as $3.48\pm0.24$~Mpc based on the TRGB standard candle, which makes it strikingly isolated, with no known galaxy within $\sim$1~Mpc.

Corvus~A is several times more massive than either Leo~P or Pavo and occupies the space between these objects and the lowest mass galaxies in the SHIELD sample. However, Corvus~A is considerably closer than the vast majority of the SHIELD galaxies, making it better suited to higher resolution follow-up at both radio and optical wavelengths. SEAMLESS has identified several blue candidates that likely fall in this regime and, once confirmed, will begin to populate the space between Leo~P and Pavo and the SHIELD sample. With the advent of the Rubin Observatory's Legacy Survey of Space and Time it will soon be possible to construct a statistical sample of nearby galaxies in this mass range through similar search techniques.

\begin{acknowledgments}
This work uses VLA observations from project 23B-188. The National Radio Astronomy Observatory is a facility of the National Science Foundation operated under cooperative agreement by Associated Universities, Inc.
This paper includes data gathered with the 6.5 meter Magellan Telescopes located at Las Campanas Observatory, Chile. 
This work used images from the Dark Energy Camera Legacy Survey (DECaLS; Proposal ID 2014B-0404; PIs: David Schlegel and Arjun Dey). Full acknowledgment at \url{https://www.legacysurvey.org/acknowledgment/}. 
This publication uses data generated via the \url{Zooniverse.org} platform, development of which is funded by generous support, including a Global Impact Award from Google, and by a grant from the Alfred P. Sloan Foundation.
This paper uses data products produced by the OIR Telescope Data Center, supported by the Smithsonian Astrophysical Observatory. Partially based on observations obtained at the Southern Astrophysical Research (SOAR) telescope, which is a joint project of the Minist\'{e}rio da Ci\^{e}ncia, Tecnologia e Inova\c{c}\~{o}es (MCTI/LNA) do Brasil, the US National Science Foundation’s NOIRLab, the University of North Carolina at Chapel Hill (UNC), and Michigan State University (MSU).
DJS acknowledges support from NSF grants AST-1821967, 1813708 and AST-2205863.
Research by DC is supported by NSF grant AST-1814208.
KS acknowledges support from the Natural Sciences and Engineering Research Council of Canada (NSERC).
AK acknowledges support from NSERC, the University of Toronto Arts \& Science Postdoctoral Fellowship program, and the Dunlap Institute.
DZ and RD acknowledge support from NSF AST-2006785 and NASA ADAP 80NSSC23K0471 for their work on the SMUDGes pipeline.
\end{acknowledgments}

%

\vspace{5mm}
\facilities{Blanco, Magellan:Clay (Megacam), Swift, VLA, SO:Kuiper, SOAR}


\software{\href{http://astrometry.net/}{\texttt{astrometry.net}} \citep{astrometry}, \href{https://www.astromatic.net/software/scamp/}{\texttt{SCAMP}} \citep{scamp}, \href{https://www.astromatic.net/software/swarp/}{\texttt{SWarp}} \citep{swarp}, \href{https://www.astropy.org/index.html}{\texttt{astropy}} \citep{astropy2013,astropy2018}, \href{https://photutils.readthedocs.io/en/stable/}{\texttt{Photutils}} \citep{photutils}, \href{https://reproject.readthedocs.io/en/stable/}{\texttt{reproject}} \citep{reproject}, \href{https://matplotlib.org/}{\texttt{matplotlib}} \citep{matplotlib}, \href{https://numpy.org/}{\texttt{numpy}} \citep{numpy}, \href{https://scipy.org/}{\texttt{scipy}} \citep{scipy1,scipy2}, \href{https://pandas.pydata.org/}{\texttt{pandas}} \citep{pandas1,pandas2}, \href{https://astroquery.readthedocs.io/en/latest/}{\texttt{astroquery}} \citep{astroquery}, \href{https://astroalign.quatrope.org/en/latest/}{\texttt{astroalign}} \citep{astroalign}, \href{https://ccdproc.readthedocs.io/en/latest/}{ccdproc} \citep{ccdproc}, \href{https://sites.google.com/cfa.harvard.edu/saoimageds9}{\texttt{DS9}} \citep{DS9}, \href{https://users.obs.carnegiescience.edu/peng/work/galfit/galfit.html}{\texttt{GALFIT}} \citep{Peng+2002,Peng+2010}, \texttt{daophot} and \texttt{allframe} \citep{Stetson87,Stetson94}}


\appendix

\section{Region to select star for the TRGB measurement} \label{sec:aperture}

\begin{figure}
    \centering
    \includegraphics[width=0.5\textwidth]{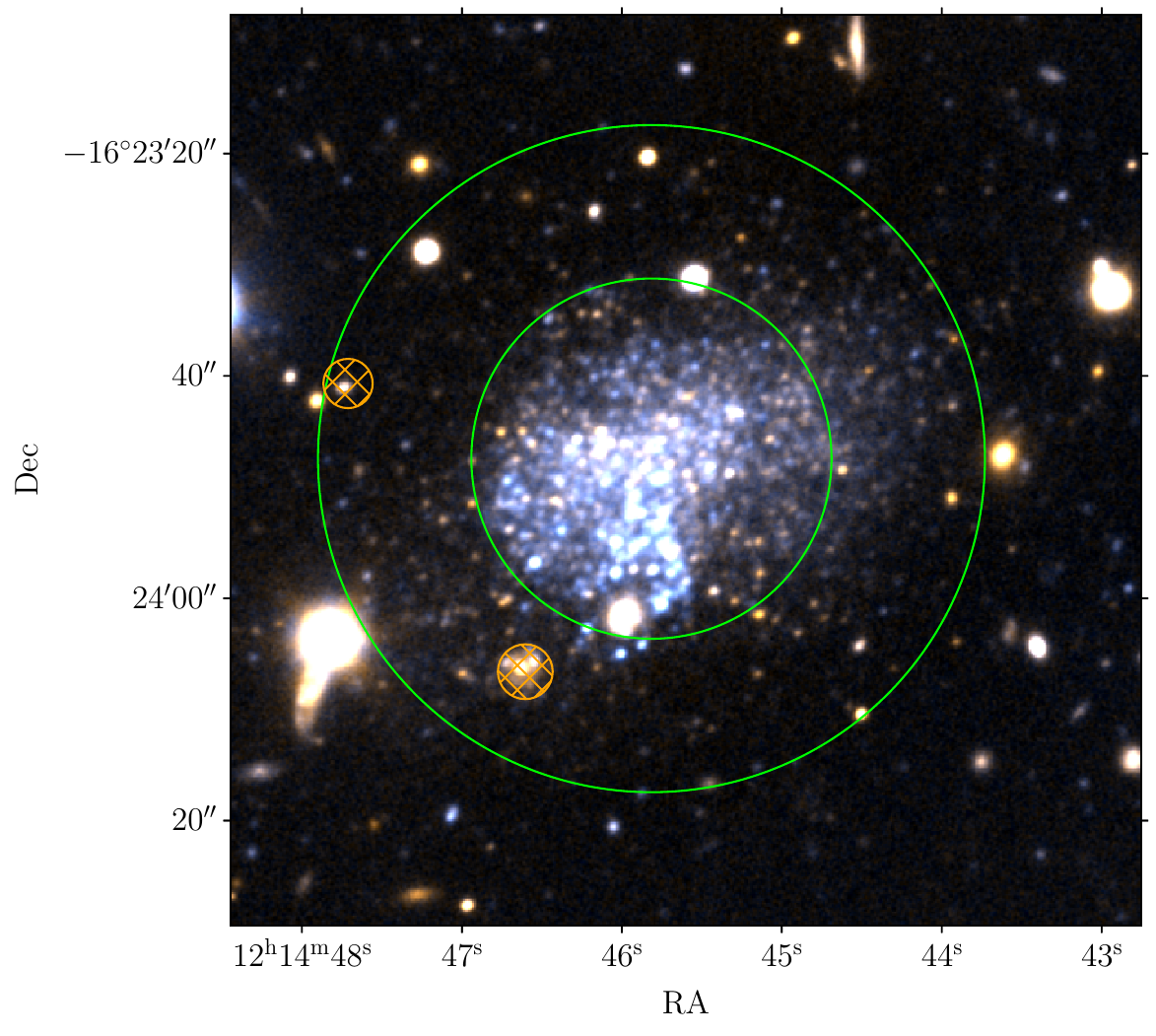}
    \caption{Magellan/Megacam $g+r$ image of Corvus~A (as in Figure~\ref{fig:Megacam}) with the annular aperture used to select stars for determining the TRGB overlaid. Bright contaminants with associated spurious point source detections were masked (orange hatched circles).}
    \label{fig:aperture}
\end{figure}

In Figure~\ref{fig:aperture} (right) we show the annular aperture (and excluded regions) used to select stars for determining the TRGB. The larger circle indicates the region used to populate the full Corvus~A CMD in Figure~\ref{fig:CMD} (left).

\bibliography{refs}{}
\bibliographystyle{aasjournal}



\end{document}